\documentclass[pra,superscriptaddress,longbibliography,twocolumn]{revtex4-2}
\usepackage[utf8]{inputenc}
\usepackage[english]{babel}
\usepackage{amsmath}
\usepackage{amsfonts}
\usepackage{amssymb}
\usepackage{array}
\usepackage{graphicx}
\usepackage{amsthm}
\usepackage{color}
\usepackage{hyperref}
\usepackage[caption=false]{subfig}
\usepackage{graphicx,float}
\usepackage{dcolumn}
\usepackage{bm}
\usepackage{mathrsfs}
\usepackage{txfonts}
\usepackage{CJK}
\usepackage[amssymb]{SIunits}
\usepackage{epsfig}
\usepackage{epstopdf}
\usepackage{lipsum}
\usepackage{placeins}
\usepackage{float}

\allowdisplaybreaks[2]

\begin{document}
\title{Dynamic Nonreciprocity with a Kerr Nonlinear Resonator}

 \author{Rui-Kai Pan}
 \affiliation{College of Engineering and Applied Sciences, National Laboratory of Solid State Microstructures, and  Collaborative Innovation Center of Advanced Microstructures, Nanjing University, Nanjing 210093, China}

 \author{Lei Tang}
 \email{tanglei@sicnu.edu.cn}
 \affiliation{College of Engineering and Applied Sciences, National Laboratory of Solid State Microstructures, and  Collaborative Innovation Center of Advanced Microstructures, Nanjing University, Nanjing 210093, China}
 \affiliation{College of Physics and Electronic Engineering, Sichuan Normal University, Chengdu 610101, China}

 \author{Keyu Xia}
 \email{keyu.xia@nju.edu.cn}
 \affiliation{College of Engineering and Applied Sciences, National Laboratory of Solid State Microstructures, and  Collaborative Innovation Center of Advanced Microstructures, Nanjing University, Nanjing 210093, China}
 \affiliation{Jiangsu Key Laboratory of Artificial Functional Materials, Nanjing University, Nanjing 210023, China}
 \affiliation{Shishan Laboratory, Suzhou Campus of Nanjing University, Suzhou 215000, China}

 \author{Franco Nori}
 \affiliation{RIKEN Quantum Computing Center, RIKEN Cluster for Pioneering Research, Wako-shi, Saitama 351-0198, Japan}
 \affiliation{Physics Department, The University of Michigan, Ann Arbor, Michigan 48109-1040, USA}

\begin{abstract}
On-chip optical nonreciprocal devices are vital components for integrated photonic systems and scalable quantum information processing. Nonlinear optical isolators and circulators have attracted considerable attention because of their fundamental interest and their important advantages in integrated photonic circuits. However, optical nonreciprocal devices based on Kerr or Kerr-like nonlinearity are subject to dynamical reciprocity when the forward and backward signals coexist simultaneously in a nonlinear system. Here, we theoretically propose a method for realizing on-chip nonlinear isolators and circulators with dynamic nonreciprocity. Dynamic nonreciprocity is achieved via the chiral modulation on the resonance frequency due to coexisting self- and cross-Kerr nonlinearities in an optical ring resonator. This work showing dynamic nonreciprocity with a Kerr nonlinear resonator can be an essential step toward integrated optical isolation.
\end{abstract}
\maketitle

\section{introduction}
 %
Optical nonreciprocal components, such as optical isolators and circulators, can isolate detrimental backscattering fields from the signal source and thus are vital for photon-based information processing in both the classical and quantum regimes~\cite{PhysRevLett.78.3221,kimble_quantum_2008,doi:10.1126/science.abe3150,PhysRevA.97.062318,Nat.Photonics.14.345}. Their realization relies on the breakdown of the Lorentz reciprocity. The magneto-optical effect~\cite{oqe1978.10,ol1986.11} is commonly used to realize optical isolators and circulators, but it is limited to bulk optics. A magnonic system using a yttrium iron garnet has been proposed to tackle this problem for realizing microwave nonreciprocity~\cite{PhysRevLett.123.127202,PhysRevA.105.013711}. The spatiotemporal modulation of optical systems has successfully demonstrated the capability of achieving optical nonreciprocity~\cite{prl2012.109, nat.phys2014.10}. A chiral atom-cavity system with spin-momentum locking can exhibit quantum nonreciprocity~\cite{pra2014.90,doi:10.1126/science.1254699,sollner_deterministic_2015,PhysRevX.5.041036,doi:10.1126/science.aaj2118,pra2019.99,PhysRevLett.128.203602,Lu:22,Science.348,Nat.Commun.10,Nat.Phys.12,ACS.Photonics.4,Nat.Photonics.9,Phys.Rep.592}. An ensemble of hot atoms provides a useful platform to achieve all-optical isolation when unidirectional control fields are used to induce susceptibility-momentum locking in atoms~\cite{nat.photon2018.12, prl2018.121, PhysRevLett.123.033902, prl2020.125, prr2020.2.033517, sci.adv2021.7.eabe8924, prapplied2021.16.014046, hu_noiseless_2021, LPR.16.2100708}. Susceptibility-momentum locking has become a new toolkit for realizing quantum nonreciprocity~\cite{https://doi.org/10.1002/qute.202200014}. By using the macroscopic Doppler shift in a unidirectional moving atomic lattice, all-optical isolators and unidirectional reflectionless have been studied~\cite{PhysRevLett.110.093901, PhysRevLett.110.223602, PhysRevLett.113.123004}. Alternatively, spinning resonators~\cite{PhysRevA.103.053522,Nature.558,PhysRevApplied.10.064037} and optomechanical resonators~\cite{nat.photon2016.10,PhysRevA.98.063845, PhysRevA.102.011502, ruesink_nonreciprocity_2016,Nature.537.80,Nature.568.65} are also be used to achieve optical nonreciprocity, in particular, magnetic-free optical isolation. In spite of the significant progress in non-magnetic optical nonreciprocal devices, the realization of integrated all-optical isolation on a solid-state platform remains very challenging.

Nonlinearity in solid-state optical materials like silicon was always a promising candidate for breaking the Lorentz reciprocity without magnetic fields or complicated spatiotemporal modulation~\cite{np2013.7.579}. Thus, nonlinear devices had attracted intense study for the realization of integrated isolators and circulators because they can be integrated on a chip with silicon-based materials and is bias-free~\cite{science2012.335, laser.photon.rev2015.9, nat.photon2020.14,PRJ2021.9.1218}.
However, an optical nonreciprocal device based on Kerr or Kerr-like nonlinearity is subject to dynamic reciprocity, which was derived from a nonlinear Helmholtz equation only including the cross-Kerr nonlinearity of the material but excluding the self-Kerr nonlinearity~\cite{nat.photon2015.9}. Because of dynamic reciprocity, this type of nonlinear nonreciprocal devices cannot work as optical isolators when the forward and backward fields coexist~\cite{science2012.335, laser.photon.rev2015.9, prb2018.97, nat.photon2020.14,PRJ2021.9.1218}.

The problem of dynamic reciprocity is the main challenge of using a nonlinear platform for integrated nonlinear optical isolators. Other types of nonlinear optical isolators using quantum nonlinearity~\cite{prl2019.123.233604} or nonlinearity in a parity-time-symmetry-broken system~\cite{nat.phys2014.10.394,nat.photon2014.8} are also subject to dynamic reciprocity. Therefore, bypassing dynamic reciprocity with a nonlinear optical device becomes highly desirable for integrated optical isolation. Dynamic reciprocity has been widely accepted as a basic knowledge in nonlinear optics.
To avoid the problem of dynamic reciprocity, pulsed signals are used in nonlinear optical isolators. In this way, the opposite-propagating signals are temporally separate, allowing isolation of the pulsed backscattering field~\cite{nat.photon2020.14,PRJ2021.9.1218}. Nevertheless, the dynamic reciprocity still limits its application in the circumstance of  continuous signals.

Some novel mechanisms  have been proposed for bypassing dynamic reciprocity in nonlinear optical isolators. These methods exploit either chiral Kerr-type nonlinearity in atomic media~\cite{prl2018.121, prr2020.2.033517, prapplied2021.16.014046}, nonlinearity-induced spontaneous symmetry breaking~\cite{optica2018.5, PhysRevA.98.053863}, and unidirectional parametric nonlinear processes~\cite{prl2021.126.133601, PRL2022.128.083604}. Nevertheless, circumventing the problem of dynamic reciprocity in a solid-state platform with Kerr nonlinearity is still desirable. Moreover, this can change the concept of dynamic reciprocity.

In this paper, we show that a Kerr-type nonlinear microring resonator (MR) can show optical dynamic nonreciprocity. The material of this MR should include self-Kerr and cross-Kerr nonlinearity and thus is compatible with silicon. Because of the intrinsic chirality of Kerr nonlinear media, the self-Kerr modulation (SKM) and cross-Kerr modulation (XKM) on the MR resonance frequency are different and dependent on the propagation of light. As a result, nonlinear optical isolators and three-port quasi-circulators based on this chiral Kerr nonlinearities can be attained for continuous inputs simultaneously propagating in opposite directions. We also employ finite-difference time-domain (FDTD)  simulations to validate the dynamic nonreciprocity predicted by the coupled-mode theory.

\section{System and model}
Our idea for realizing dynamic reciprocity with a nonlinear MR makes use of the intrinsic chirality of a nonlinear medium. A nonlinear medium like silicon possesses self- and cross-Kerr nonlinearities simultaneously. The cross-Kerr nonlinearity strength is typically twice of the self-Kerr nonlinearity~\cite{Boydbook,Xiao:08,PhysRevLett.118.033901, del_bino_symmetry_2017,optica2018.5}. Thus, if we design a system where the forward and backward light fields in a nonlinear medium are different in power, then the opposite-propagating fields will ``see'' different refractive indices of the same medium. For an optical system sensitive to the refractive index of the medium, the opposite-input fields will have different transmissions. In this way, we can realize optical isolators and circulators with dynamic reciprocity.

 \begin{figure}[!ht]
 \centering\includegraphics[width= 1\linewidth]{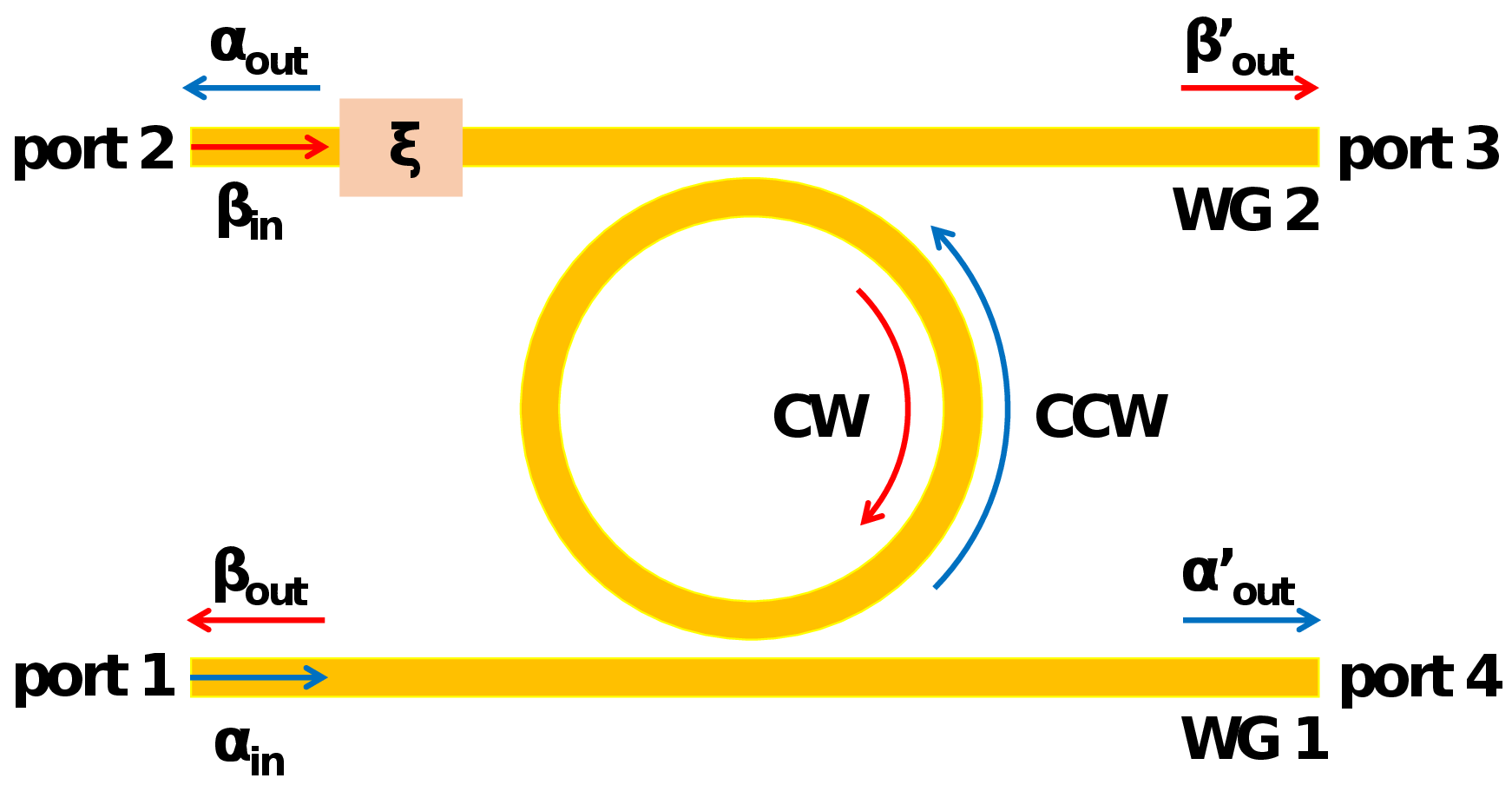}
 \caption{Schematic diagram for optical isolation using a Kerr nonlinear medium. A nonlinear microring resonator (MR) couples to the bus (WG 1) and drop (WG 2) waveguides. An optical attenuator with amplitude transmission $\xi$ is embedded in the left end of the WG 2.  The incident light $\alpha_\text{in}$ ($\beta_\text{in}$) from port 1 (port 2) respectively excites the CCW (CW) mode of the MR.}
 \label{fig:system}
 \end{figure}

The system consists of a Kerr-nonlinear MR, a bus waveguide (WG 1) with ports 1 and 4, and a drop waveguide (WG 2) with ports 2 and 3, as shown in Fig.\,1.
An optical attenuator with amplitude transmission $\xi$ is embedded near port 2 inside the WG 2.
A continuous forward light with power $P_\text{in}$ and frequency $\omega_\text{in}$ inputs to port 1 and then excites the counterclockwise (CCW) mode in the MR. The corresponding photon flux is $|\alpha_\text{in}|^2=P_\text{in}/\hbar\omega_\text{in}$. To test the functionality of the proposed nonlinear optical circulator, a continuous backward light field with the same power $P_\text{in}$, frequency and photon flux $|\beta_\text{in}|^2=P_\text{in}/\hbar\omega_\text{in}$ simultaneously drives the clockwise (CW) mode from port 2.
In the frame rotating at frequency $\omega_\text{in}$, the Hamiltonian of the system with both the self- and cross-Kerr nonlinearities is given by~\cite{PhysRevLett.118.033901,Xiao:08}
 \begin{equation}\label{eq:H}
 \begin{aligned}
  H = & \Delta a^{\dagger} a + \Delta b^{\dagger} b + U a^{\dagger 2} a^2  + U b^{\dagger 2}b^2 + 4U a^{\dagger}a b^{\dagger} b \\
  & + i \sqrt{2\kappa_{\text{ex1}}}(\alpha_{\text{in}} a^{\dagger} - \alpha_{\text{in}}^* a) + i \xi\sqrt{2\kappa_{\text{ex2}}}(\beta_{\text{in}} b^{\dagger}- \beta_{\text{in}}^* b)\;,
  \end{aligned}
 \end{equation}
where $\Delta = \omega_0 - \omega_\text{in}$, $\omega_0$ is the resonance frequency of the MR,  $a$ ($b$) represents the CW (CCW) mode of the MR, $\kappa_{\text{ex1}}~(\kappa_{\text{ex2}})$ is the external decay rate due to the WG 1-resonator (WG
2-resonator) coupling, $\kappa_{\text{i}}$ is the intrinsic decay rate of the MR. Thus, $\kappa = \kappa_{\text{ex1}} + \kappa_{\text{ex2}} +\kappa_{\text{i}}$ is the total decay rate of the MR.
$U$ is the Kerr nonlinearity strength, given by $U = \hbar \omega_0^2 c n_2 / (n_0^2 V_\text{m})$~\cite{marin-palomo_microresonator-based_2017}, where $n_0$ and $n_2$ are the linear and nonlinear refraction indices of the medium, respectively, and $V_{\text{m}}$ is the effective mode volume of the MR.
The third and fourth terms model the SKM, whereas the sixth term describes the XKM.

The device is an optical isolator when we only consider ports 1 and 2 as the input and output ports. If the port 3 is included, it can work as a three-port optical quasi-circulator.

\section{Coupled-mode method}

Using Eq.~\eqref{eq:H}, the coupled-mode equation can be written as
\begin{subequations}
\begin{align}
   \dot{a} = & -(i\Delta+\kappa)a - 4iU|b|^2 a
  - 2iU|a|^2 a + \sqrt{2\kappa_{\text{ex1}}}\alpha_{\text{in}}\;,\label{seq:coupled-mode1}\\
  \dot{b} = & -(i\Delta+\kappa)b - 4iU|a|^2 b
  -2iU|b|^2 b + \xi\sqrt{2\kappa_{\text{ex2}}}\beta_{\text{in}}\;.\label{seq:coupled-mode2}
 \end{align}
  \label{eq:coupled-mode}
\end{subequations}
We solve Eq.~\eqref{eq:coupled-mode} and find the coherent amplitudes of the resonator modes, i.e., $\alpha=\langle a \rangle$ and $\beta=\langle b \rangle$.
Note that the XKM strength is twice of the SKM according to Eq.~\eqref{eq:coupled-mode}. This means that the light in the CCW mode can generate a frequency modulation to the CW mode as twice as that caused by itself with the same power. Thus, if opposite-input light fields with the same power can excite the CW and CCW mode to different power levels, by introducing an attenuator to one input port, then the two input fields ``see'' resonator modes with different resonance frequencies and thus have different transmission. \emph{This is the key idea of our optical isolator.} By including both the self- and cross-Kerr nonlinearities with different strengths, our system is crucially different from that for demonstrating dynamic reciprocity. Therefore, our system can achieve \emph{dynamic nonreciprocity}, \emph{allowing to implement optical isolators and circulators.}

According to the input-output relationship, we have the outputs
\begin{subequations}
\begin{align}
  \alpha_{\text{out}} = &~\xi \sqrt{2\kappa_{\text{ex2}}}\alpha\;, \quad \alpha_{\text{out}}' = ~\alpha_{\text{in}} - \sqrt{2\kappa_{\text{ex1}}}\alpha\;,\\
  \beta_{\text{out}} = &~\sqrt{2\kappa_{\text{ex1}}}\beta\;, \quad
  \beta_{\text{out}}' = ~\xi  \beta_{\text{in}} - \sqrt{2\kappa_{\text{ex2}}}\beta\;.
\end{align}
  \label{eq:output}
\end{subequations}
The transmissions can be calculated as
\begin{equation}
  T_{\text{12}} = \frac{|\alpha_\text{out}|^2} {|\alpha_\text{in}|^2} \;,
  T_{\text{21}} = \frac{|\beta_\text{out}|^2} {|\beta_\text{in}|^2} \;,
  T_{\text{14}} = \frac{|\alpha_\text{out}^\prime|^2} {|\alpha_\text{in}|^2} \;,
  T_{\text{23}} = \frac{|\beta_\text{out}^\prime|^2} {|\beta_\text{in}|^2} \;.
\label{eq:transmission}
\end{equation}
In calculation of $T_{ij}$, $i$ is for the input port and $j$ for the output port.

This work aims to show an optical isolator and a three-port quasi-circulator~\cite{doi:10.1126/science.aaj2118,adpr.202000104,PRL2022.128.083604} with dynamic nonreciprocity.  Thus, we are interested in the transmissions $T_{12}$, $T_{21}$ and $T_{23}$.

For the isolator, the isolation contrast and the insertion loss are defined as
\begin{equation}
  \eta = \frac{T_{12} -T_{21}}{T_{12}+T_{21}} \;,
  \label{eq:eta}
\end{equation}
and
\begin{equation}
  \mathscr{L}=-10\log_{10}(T_{12})\;,
\end{equation}
respectively. For the quasi-circulator, the average fidelity is given by
\begin{equation}
  \mathcal{F} = \frac{\text{Tr} \left[ \tilde{T} T_\text{id}^T \right]}{\text{Tr} \left[ T_\text{id} T_\text{id}^T \right]}\;,
  \label{fidelity}
\end{equation}
where $T_\text{id}$ is the transmission matrix for an idea three-port quasi-circulator and $T_\text{id} = [0~ 1 ~ 0; 0~ 0~ 1]$ as in~\cite{doi:10.1126/science.aaj2118,prl2018.121,adpr.202000104,PRL2022.128.083604}, and
\begin{equation}
  \tilde{T} = \frac{T_{ij}}{\Upsilon_i}\;,
  \label{eq:Tcirculator}
\end{equation}
 with $\Upsilon_i = \textstyle \sum_{j}T_{ij}$. The average insertion loss is defined as
 \begin{equation}
 \tilde{\mathscr{L}} = -10\log_{10}[(T_{12} + T_{23})/2]\;.
 \label{eq:insertionloss}
 \end{equation}
 Below, for simplicity, we will name a quasi-circulator as a circulator.

\section{Finite-Difference Time-Domain simulations}

As a first-principle method, FDTD simulations can predict the behavior of an electromagnetic system very precisely~\cite{FDTDbook}.
In this work, we also employ FDTD simulations to verify the results obtained from the coupled-mode theory. To reduce the calculation time, we performed two-dimensional FDTD simulations which do not include the $z$ dimension. The perfect match layer boundary condition is applied for surrounding the simulation region. The cross-section of the MR and two waveguides have the same width of $200~\text{nm}$. The radius of the MR is assumed to be $700~\text{nm}$. Taking silicon as the nonlinear medium, the refractive index of the MR and two waveguides is $n = 3.478$. The third-order nonlinear susceptibility of the nonlinear medium of the MR is taken to be $\chi^{(3)} = 2.8 \times 10^{-18}~\rm m^2/V^2$, corresponding to a nonlinear refractive index $n_2 = 2.7 \times 10^{-14}~\centi\meter^2/\watt$~\cite{Boydbook}. The optical attenuator in WG 2 has an amplitude transmission $\xi \approx 0.982$. The gaps between the MR and the two waveguides are the same, $250~\nano\meter$ so that $\kappa_{\text{ex1}} = \kappa_{\text{ex2}}~\textgreater~\kappa_{\text{in}}$. To avoid strong backscattering, the spatial grid size for the simulation is set to be small enough, $5~\nano\meter$. The temporal step of the simulation is $0.012~\femto\second$ and the total simulation time is $50~\pico\second$, allowing the system to evolve to its steady state. To show dynamic nonreciprocity, two constant driving fields are incident to ports 1 and 2 at the same time. The input light fields $E$ have the same wavelength of $\lambda = 1.685 ~\micro\meter$ and are equal.

\section{Results}
\subsection{Time-dependent Transmission}

Because the response of the system is a nonlinear function of the power of the input field, it is difficult to find an analytical steady-state solution for the transmission. Thus, we first study the time-evolution of the system.
The time-dependent transmissions are found by solving the coupled-mode equation and also via FDTD numerical simulations, as shown in Fig.\,2. The former is computation-resource-efficient and can show the underlying physics. It can also find the solution very fast. The later provides first-principle simulations for a real device. The FDTD simulations results are in reasonable agreement with the coupled-mode theory even in the details of the oscillation parts.

When solving the coupled-mode equation, two constant drivings are applied at the same time. We take $\alpha_\text{in} = \beta_\text{in} = 50\sqrt{\kappa}$ as an example. The transmissions reach their steady-state values after some oscillation over a period of about $10 \kappa t$. The steady-state transmissions are $T_{12} \approx 0.74 $, $T_{21} \approx 0.04$, $T_{23} \approx 0.91$ and $T_{14} \approx 0.06$, respectively. Obviously, the forward transmission $T_{12}$ is much higher than the backward transmission $T_{21}$. Thus, an optical isolator with ports 1 and 2 is obtained. The isolation constant is $\eta \approx 0.90$. The insertion loss of the transparent direction is low, $\mathscr{L} \approx 1.31 ~\text{dB}$. The telecom-wavelength signal can also transmit along the direction $1\rightarrow2\rightarrow3$, forming a three-port circulator with $\mathcal{F}\approx 0.98$ and $\tilde{\mathscr{L}} \approx 0.83~\text{dB}$. \emph{The results of the couple-mode theory clearly show an optical isolator and a quasi-circulator with high dynamic nonreciprocity and low insertion loss.}

\begin{figure}[!ht]
\centering\includegraphics[width= 1\linewidth]{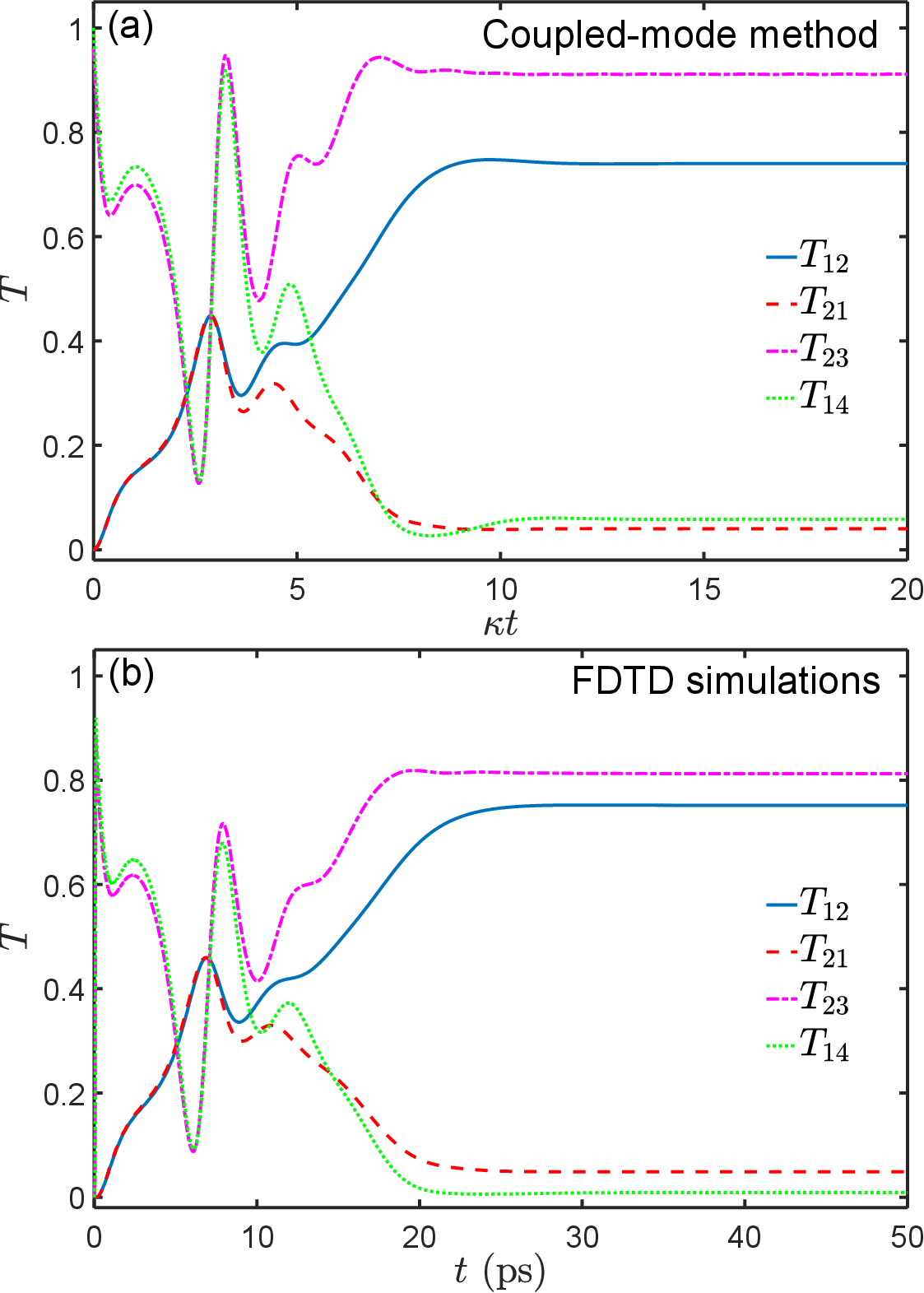}
\caption{Time-dependent nonreciprocal transmission. (a) transmission predicted by the coupled mode equations with parameters: $\kappa_{\text{ex1}} = \kappa_{\text{ex2}} = 0.45\kappa$, $\kappa_{\text{i}} = 0.1\kappa$, $\xi = 0.98$, $\Delta = -4.5\kappa$, $U = 0.001\kappa$, and $P_{\text{in}}/{\hbar \omega_{\text{in}}} = 2500 \kappa$. (b) transmission obtained from the FDTD simulations with $E = 2.2 \times 10^7 ~\volt/\meter$.}
\label{fig:transmission_t}
\end{figure}

\begin{figure}[!ht]
\centering\includegraphics[width= 1\linewidth]{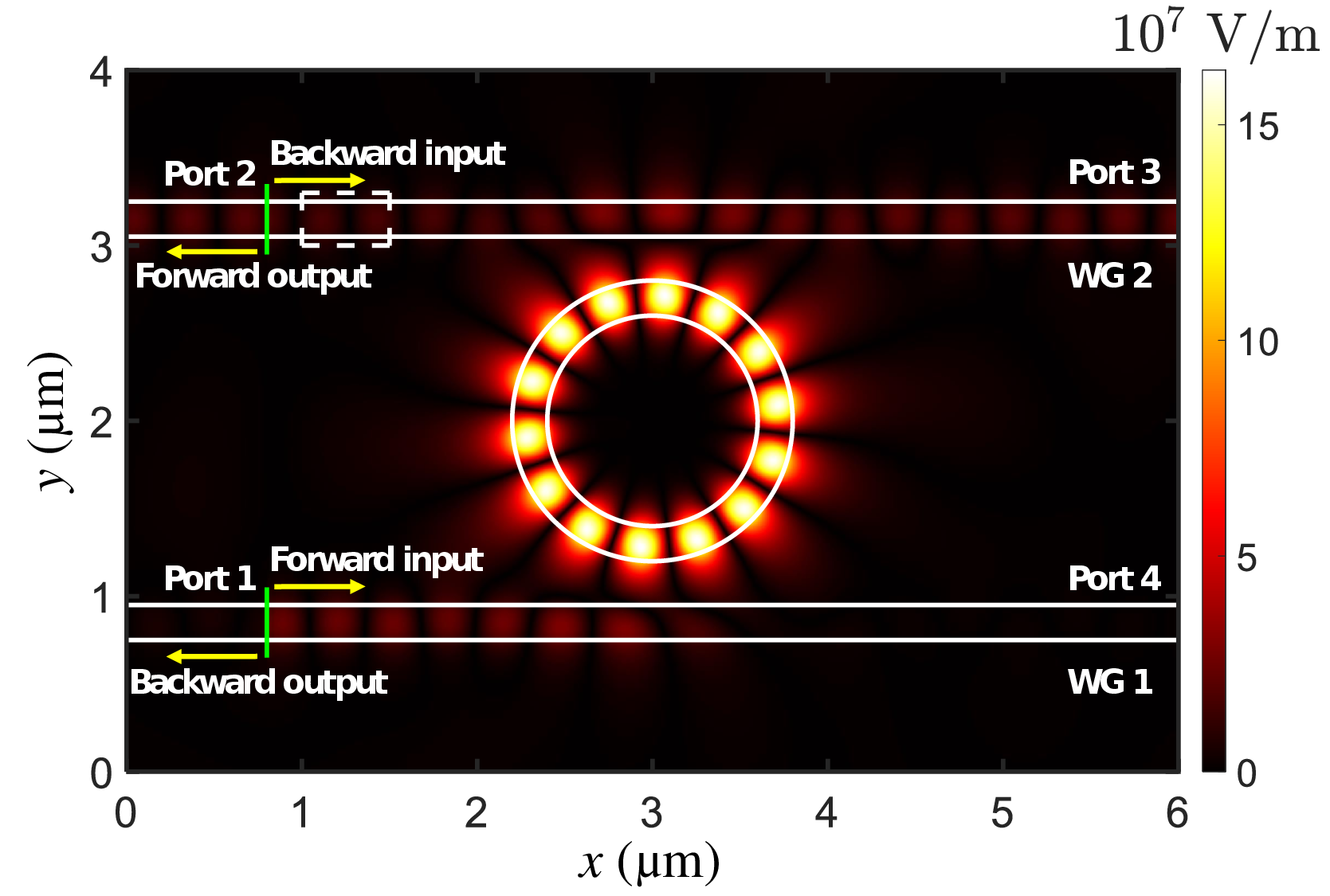}
\caption{Distribution of the instantaneous electric field at $t = 50~\pico\second$. The forward and backward continuous light is incident to ports 1 and 2 simultaneously. The position of the forward and backward sources are labeled by green vertical lines in the WG 1 and WG 2, respectively. The optical attenuator is marked by the dashed rectangle in the upper left.}
\label{fig:profile}
\end{figure}

The time evolution and steady-state values of the transmissions in the FDTD numerical simulations are very close to the results of the coupled-mode theory. By calculating the intensities of the transmitted fields in the FDTD simulations, we obtain $T_{12} \approx 0.75 $, $T_{21} \approx 0.05$, $T_{23} \approx 0.81$ and $T_{14} \approx 0.01$, respectively. These results yield the isolation contrast $\eta \approx 0.88$ and the insertion loss of $\mathscr{L}\approx1.24 ~\text{dB}$ for the two-port optical isolator and the average fidelity of $\mathcal{F} \approx 0.97$ and the average insertion loss of $\tilde{\mathscr{L}}\approx1.07~\text{dB}$ for the three-port circulator~\cite{OE.21.5041, optica2018.5,PRL2022.128.083604}. Note that the transmission $T_{23}$ is high. This high $T_{23}$ is important for optical sensors~\cite{Nat.Phys.4.472,Opt.Lett.42.290,sciadv.aaw1899,Nat.Photonics.14.345,nat.photon2020.14}. The FDTD simulation results are in good agreement with the coupled-mode theory, see Tables~1 and 2 for comparison.
There is a small discrepancy between these two methods because the structure parameters in the FDTD simulation cannot perfectly match those of the coupled-mode method.

\begin{table}[!ht]
\caption{Transmission.}
    \centering
    \begin{tabular}{|l|p{1.2cm}|p{1.2cm}|p{1.2cm}|p{1.2cm}|}
    \hline
    \hline
        ~ & $T_{12}$ & $T_{21}$ & $T_{23}$ & $T_{14}$ \\ \hline
        Coupled-mode method & 0.74 & 0.04 & 0.91& 0.06 \\ \hline
        FDTD simulation & 0.75 & 0.05 & 0.81 & 0.01 \\
    \hline
    \end{tabular}
    \label{tab:transmissions}
\end{table}

\begin{table}[!ht]
\renewcommand{\arraystretch}{1.2}
\caption{Performance.}
    \centering
    \begin{tabular}{|l|p{1.2cm}|p{1.2cm}|p{1.2cm}|p{1.2cm}|}
    \hline
    \hline
        ~ & $\eta$ & $\mathscr{L}$ &  $\mathcal{F}$ &  $\tilde{\mathscr{L}}$ \\ \hline
        Coupled-mode method & 0.90 & 1.31 dB & 0.98& 0.83 dB \\ \hline
        FDTD simulation & 0.88 & 1.24 dB & 0.97 & 1.07 dB \\
        \hline
    \end{tabular}
    \label{tab:performance}
\end{table}

Figure~\ref{fig:profile} shows the instantaneous electric field distribution at $t = 50~\pico\second$ as an example when the continuous light is simultaneously incident to ports 1 and 2. It can be seen that the forward light incident to port 1 exits from port 2 with a high transmission. However, the backward light from port 2 dominantly transmits to port 3. The transmission to port 1 is negligible.
These FDTD simulations clearly prove dynamic nonreciprocity of the nonlinear MR and confirm this design for a practical optical circulator.

\subsection{Steady-state Transmission}

\begin{figure}[!ht]
\centering\includegraphics[width=1\linewidth]{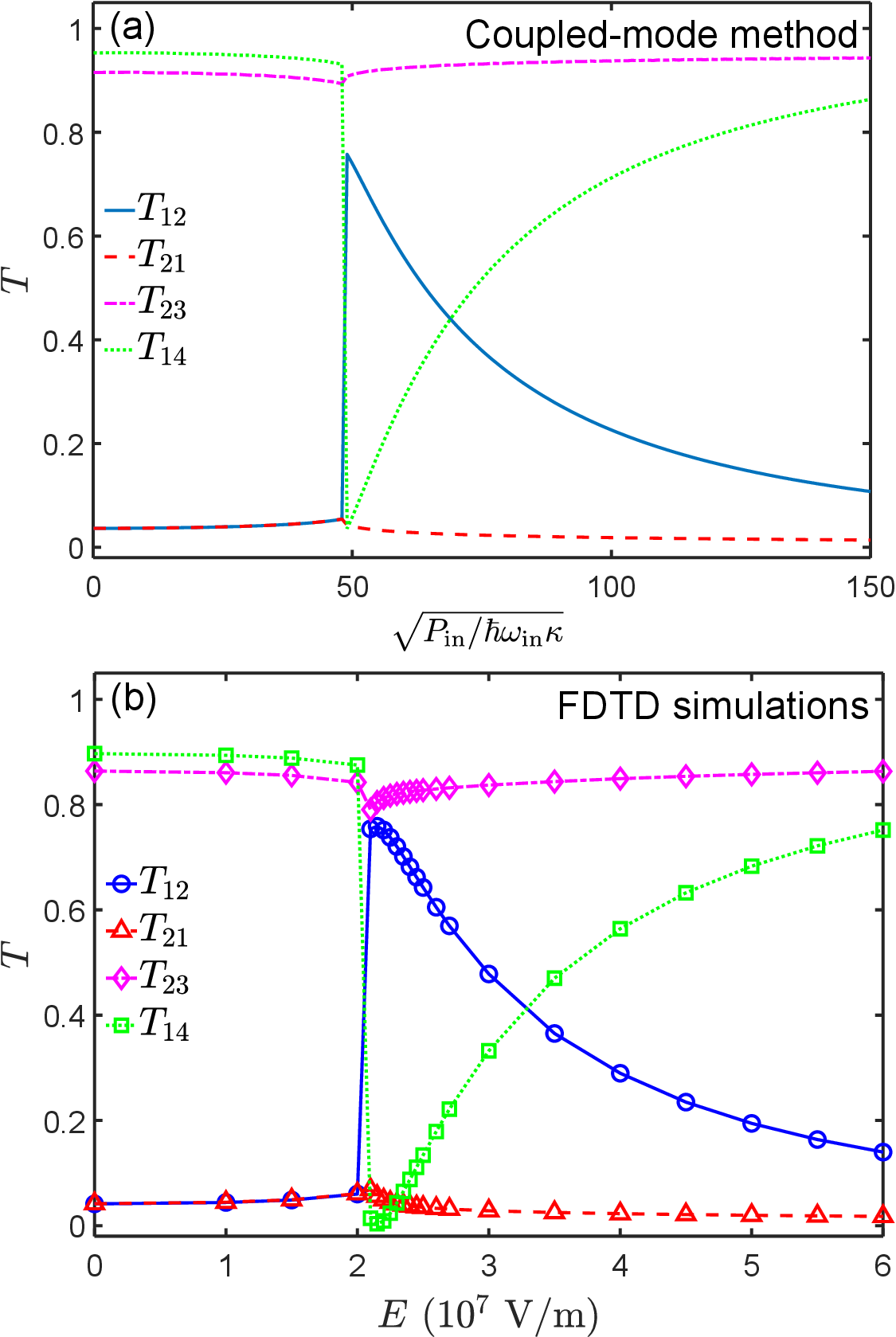}
\caption{Steady-state results versus input intensity. (a) The transmission by numerically solving the coupled-mode equations with parameters: $\kappa_{\text{ex1}} = \kappa_{\text{ex2}} = 0.45\kappa$, $\kappa_{\text{i}} = 0.1\kappa$, $\xi = 0.98$, $\Delta = -4.5\kappa$ and $U = 0.001\kappa$.  (b) Steady-state transmission according to FDTD simulations.}
\label{fig:ain}
\end{figure}

\begin{figure*}
\centering\includegraphics[width=1\linewidth]{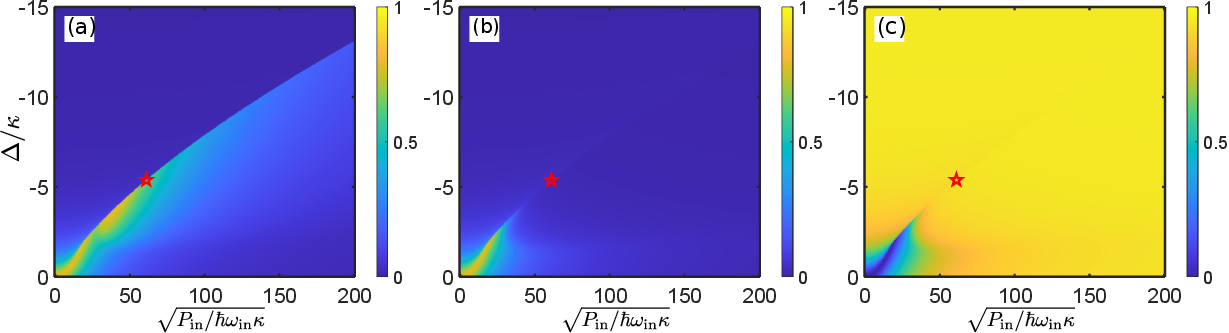}
\caption{Steady-state transmission by solving the coupled-mode equation. Transmission $T_{12}$ (a), $T_{21}$ (b) and $T_{23}$ (c) versus input intensity and detuning. The red star indicates an optimal point for a trade-off between isolation contrast and insertion loss. Other parameters are: $\kappa_{\text{ex1}} = \kappa_{\text{ex2}} = 0.45\kappa$, $\kappa_{\text{i}} = 0.1\kappa$, $\xi = 0.98$, and $U = 0.001\kappa$.}
\label{fig:aindeltaT}
\end{figure*}

\begin{figure*}
\centering\includegraphics[width=1\linewidth]{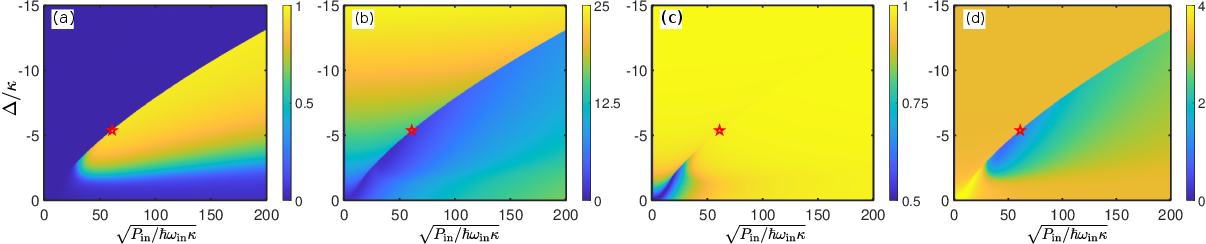}
\caption{Steady-state isolation contrast $\eta$ (a) and insertion loss $\mathscr{L}$ (b) for the optical isolator, fidelity $\mathcal{F}$ (c) and average insertion loss $\tilde{\mathscr{L}}$ (d) for the circulator versus the input intensity and detuning. The red star indicates a trade-off point between isolation performance and insertion loss. Other parameters are: $\kappa_{\text{ex1}} = \kappa_{\text{ex2}} = 0.45\kappa$, $\kappa_{\text{i}} = 0.1\kappa$, $\xi = 0.98$, and $U = 0.001\kappa$.}
\label{fig:aindeltaP}
\end{figure*}

As shown in Fig.~\ref{fig:transmission_t}, the nonlinear system can reach steady state after some time. We consider the transmissions at $\kappa t =30$ in the coupled-mode theory or $t = 50~\pico\second$ in the FDTD simulations as steady-state transmissions.
Below, we investigate the steady-state transmission versus the input power.

The transmission is dependent on the detuning between the input and the MR. Specifically, we choose $\Delta = - 4.5 \kappa$ when solving the coupled-mode equation. When $\sqrt{P_\text{in}/\hbar \omega_\text{in}} < 49 \sqrt{\kappa}$, the forward and backward transmissions are almost equal. The system is reciprocal. There is a rapid transition at $\sqrt{P_\text{in}/\hbar \omega_\text{in}} \approx 49 \sqrt{\kappa}$. We find that the forward and backward transmissions become very different after this point. For example, $T_{12} = 0.76$ and $T_{21} = 0.04$ at $\alpha_\text{in} = \beta_\text{in} = \sqrt{P_\text{in}/\hbar \omega_\text{in}} = 49 \sqrt{\kappa}$, yielding $\eta = 0.89$ and $\mathscr{L} = 1.21~\deci\bel$. As the input power increases, the transmission $T_{12}$ exponentially decreases, while $T_{21}$ remains small. When $\sqrt{P_\text{in}/\hbar \omega_\text{in}} \approx 49 \sqrt{\kappa}$, $T_{14}$ jumps to a small value from a high transmission and then increases exponentially with the input power. In contrast, $T_{23}$ remains high.

As other demonstrated nonlinear nonreciprocal devices with dynamic reciprocity~\cite{science2012.335,laser.photon.rev2015.9,optica2018.5,prb2018.97,PRJ2021.9.1218}, our system shows strong dynamic nonreciprocity only for a narrow input power
\begin{equation}
49 \sqrt{\kappa} \lesssim \sqrt{P_\text{in}/\hbar \omega_\text{in}} \lesssim 55 \sqrt{\kappa} \;,
\end{equation}
 where the insertion loss is less than $2~\deci\bel$.

Figure~\ref{fig:ain}(b) shows the results of FDTD simulation versus the input field strength $E$. The four transmissions show similar dependence on the input field $E$ as Fig.~4(a). The transition point for transmissions $T_{12}$ and $T_{14}$ is at $E \approx 2.1 \times 10^7 ~\volt\per\meter$. The transmission $T_{12}$ ($T_{14}$) jumps up (down) from a small (large) value to a large (small) value and then exponentially decreases (increase) with the input field strength. These results provide a proof of dynamical nonreciprocity predicted by the coupled-mode theory.

The response of a nonlinear system is crucially dependent on the input field parameters, such as frequency and the input power.
 Figure~\ref{fig:aindeltaT} shows the transmissions $T_{12}$, $T_{21}$ and $T_{23}$ as functions of the power and detuning of the incident light by solving the coupled-mode equations. Figures~\ref{fig:aindeltaP}(a) and ~\ref{fig:aindeltaP}(b) show the isolation contrast $\eta$ and the insertion loss $\mathscr{L}$ for the optical isolator, respectively. Figures~\ref{fig:aindeltaP}(c) and ~\ref{fig:aindeltaP}(d) show the average fidelity $\mathscr{L}$ and the average insertion loss $\tilde{\mathscr{L}}$ for the three-port circulator, respectively.
Obviously, the transmission $T_{12}$ approaches $T_{21}$ for small detuning and weak input power because the power-dependent SKM and XKM are weak and can only cause slightly different frequency shifts to the CW and CCW modes. The system shows strong dynamic nonreciprocity when the detuning and power are large enough, see Fig.~\ref{fig:aindeltaP}(a). However, when the detuning or the input power is too large, the transparent transmission is low, implying a large insertion loss, see Fig.~\ref{fig:aindeltaP}(b).

Optical isolation requires a trade-off between the isolation contrast and the insertion loss.
An optimal point is indicated by the red star in Figs.~5 and 6,
where $\sqrt{P_\text{in}/\hbar \omega_\text{in}} \approx 61 \sqrt{\kappa}$ and $\Delta \approx  -5.38 \kappa$. At this optimal point, the forward and backward transmissions of the optimal point are $T_{12} \approx 0.64$ and $T_{21} \approx 0.02$, corresponding to the isolation contrast $\eta \approx 0.94$ and an insertion loss of $\mathscr{L} \approx 1.96~\text{dB}$ for the optical isolator. Considering the transmission  $T_{23} \approx 0.93$, the system can work as a three-port circulator with the high performance of $\mathcal{F} \approx 0.99$ and $\tilde{\mathscr{L}} \approx 1.05~\text{dB}$. The transmission and performance of the optical nonreciprocal device at the optimal point are summarized in Table.~\ref{tab:transmission and performance}.

\begin{table}[!ht]
\renewcommand{\arraystretch}{1.2}
\caption{Transmission and performance.}
    \centering
    \begin{tabular}{|p{1cm}|p{1cm}|p{1cm}|p{1cm}|p{1cm}|p{1cm}|p{1cm}|}
    \hline
    \hline
         $T_{12}$ & $T_{21}$ & $T_{23}$ & $\eta$ & $\mathscr{L}$ & $\mathcal{F}$ & $\tilde{\mathscr{L}}$ \\ \hline
         0.64 & 0.02 & 0.93 & 0.93 & 1.96 dB & 0.99& 1.05 dB \\
        \hline
    \end{tabular}
    \label{tab:transmission and performance}
\end{table}

Because the system is nonlinear, the performance of the device is crucially dependent on the total input power in the microresonator, the cavity-input detuning and many other system parameters. Thus, it is difficult to find the global optimal parameters to achieve a high isolation ratio for such a nonlinear system.
Here, we aim to show an optical nonreciprocal device with a small insertion loss for two input fields with \emph{very close powers in opposite directions}, namely, $\xi = 0.98$ and $|\alpha_\text{in}| = |\beta_\text{in}|$. The attenuator only induces a very small absorption to the transmitted field in the forward case and to the input field to port 2 in the backward case. In this arrangement, the input powers of light entering the microresonator are very close in the forward and backward cases. As a result, the difference of the frequency shift due to the SKM and XKM is small and the isolation ratio is not very high. Nevertheless, our device using a nonlinear microring resonator exhibits the dynamic nonreciprocity with a low insertion loss and a usable isolation contrast.

%
\section{Implementation}
Now we discuss the experimental implementation of our proposal. Our scheme can be implemented with MRs made from high-$\chi^{(3)}$ nonlinear materials, such as potassium titanyl phosphate~\cite{Zielinska:17}, Si~\cite{Tsang_2008}, SiC~\cite{Guo:21}, InP~\cite{Hong_Li_2007}. We assume the resonance frequency of the nonlinear MR $\omega_0/2\pi=193.6~\text{THz}$ and the intrinsic quality factor $Q_i=1\times 10^7$~\cite{AQS.2.041702,NP.12.297,Optica.6.380,Optica.4.1536}, which is already available in experiments. The intrinsic loss of the resonator is calculated to be about $\kappa_i \approx 2\pi \times 19.4~\text{MHz}$, and thus the total loss rate is about $\kappa = 10 \kappa_i  \approx 2\pi \times 0.194~\text{GHz}$. We select experimentally accessible parameters: $n_0 = 1.4$, $n_2 = 5.1 \times 10^{-15}~\rm m^2/W$~\cite{Zielinska:17,PhysRevA.97.013843,PhysRevLett.121.153601}, $V_\text{m} = 100~\rm \mu m^3$~\cite{spillane_ultrahigh-q_2005} and thus the nonlinearity strength is calculated to be $U \approx 2\pi \times 0.194~\text{MHz} \sim 0.001\kappa$. Using such nonlinear MR, we can achieve an optical isolator for parameters $\{\sqrt{P_{\text{in}}/{\hbar \omega_{\text{in}}}}=50\sqrt{\kappa}, \Delta=-4.5\kappa\}$ and $\{\sqrt{P_{\text{in}}/{\hbar \omega_{\text{in}}}}= 61\sqrt{\kappa},\Delta=-5.38\kappa\}$ corresponding to an input power $P_\text{in}\approx0.39~\micro\watt$ and $P_\text{in}\approx0.58~\micro\watt$, respectively.

 A high Q factor can amplify the nonlinear effect in a microresonator, and thus is crucial for achieving a high-performance optical nonreciprocal device. According to our investigation with the coupled-mode method, the nonlinear system shows weak nonreciprocity when the intrinsic Q factor is not high enough. A high Q factor can cause a larger isolation contrast and a lower insertion loss for given input powers and system parameters. We evaluate the performance of our optical nonreciprocal device with parameters scaled by the intrinsic decay rate of the microresonator. For instance, a device with a relative low intrinsic Q factor $10^5$ can still achieve almost the same performance as that using $Q = 10^7$ and $\sqrt{P_\text{in}/\hbar\omega_\text{in}} = 50 \sqrt{\kappa}$,  when $\sqrt{P_\text{in}/\hbar\omega_\text{in}} = 500 \sqrt{\kappa}$ and $U = 10^{-5} \kappa$ but keeping the ratios of other parameters with respect to $\kappa$. Here, we still keep $\kappa_\text{ex1} = 0.45 \kappa$, $\kappa_\text{i} = 0.1\kappa$ and $\kappa_\text{ex2} = 0.45\kappa$.

The difference of SFM and XFM in Kerr-type nonlinear material has been exploited to demonstrate optical isolators and circulators, but only for very different opposite input powers~\cite{optica2018.5,arXiv2206.01173}. In the work~\cite{optica2018.5}, the backward light is $3.3~\deci\bel$ and $5~\deci\bel$ lower than the forward signal.
In applications of optical sensors~\cite{Nat.Phys.4.472,Opt.Lett.42.290,sciadv.aaw1899,Nat.Photonics.14.345,nat.photon2020.14}, the backward signal has a power very close to the forward one. Thus, our nonlinear device is suitable for optical sensing.
Note that an experiment~\cite{optica2018.5} has demonstrated a passive nonlinear optical isolator by exploiting the different SKM and XKM. Very recently, another experimental group~\cite{arXiv2206.01173} has also achieved optical nonreciprocity on a chip with the same idea. However, these two works use different configurations with \emph{very different} opposite input powers. In sharp contrast to this, our system exhibits optical nonreciprocal transmission for the \emph{same} input powers in opposite directions, namely, $|\alpha_{in}|^2 = |\beta_{in}|^2$ in our work. Nevertheless, these experiments provide strong support to our proposal based on the chirality of the SKM and XKM in a Kerr-type nonlinear medium.


\section{Conclusion}
In conclusion, we have proposed a method to realize nonlinear optical isolators and circulators based on the chirality of SKM and XKM in the nonlinear MR. We have proved dynamic nonreciprocity of this nonreciprocal device with both the coupled-mode theory and FDTD simulations. The proposed scheme paves the way to realize on-chip optical isolation, and thus can boost the integration of photonic chips.

\section{Acknowledgements}
This work was supported by the National Key R\&D Program of China (Grants No. 2017YFA0303703, No. 2019YFA0308700), the National Natural Science Foundation of China (Grant Nos. 11874212, 11890704, 11690031), the Fundamental Research Funds for the Central Universities (Grant No. 021314380095), the Program for Innovative Talents and Entrepreneurs in Jiangsu, and the Excellent Research Program of Nanjing University (Grant No. ZYJH002). F.N. is supported in part by:
Nippon Telegraph and Telephone Corporation (NTT) Research, the Japan Science and Technology Agency (JST) (via the Quantum Leap Flagship Program (Q-LEAP), and the Moonshot R\&D Grant Number JPMJMS2061), the Japan Society for the Promotion of Science (JSPS) (via the Grants-in-Aid for Scientific Research (KAKENHI) Grant No. JP20H00134), the Army Research Office (ARO) (Grant No. W911NF-18-1-0358), the Asian Office of Aerospace Research and Development (AOARD) (via Grant No. FA2386-20-1-4069), and the Foundational Questions Institute Fund (FQXi) via Grant No. FQXi-IAF19-06.


%

\end{document}